\documentclass[prl,amsmath,amssymb,superscriptaddress,floatfix,nofootinbib,twocolumn]{revtex4-1}
\usepackage[utf8]{inputenc}
\usepackage{mathrsfs,bm}
\usepackage{natbib}
\usepackage{ulem}
\usepackage{txfonts}
\usepackage{amssymb}
\usepackage{rotating}
\usepackage{indentfirst}
\usepackage{graphicx,booktabs}
\usepackage{multirow}
\usepackage{slashed}
\usepackage{overpic}
\usepackage{color}
\usepackage{xcolor}
\usepackage{amssymb}
\usepackage{hyperref}
\usepackage{bbding}
\usepackage{url}

\begin{document}

\title{Implications of new evidence for lepton-universality violation in $b\to s\ell^+\ell^-$ decays}

\author{Li-Sheng Geng}
\affiliation{School of Physics \& Beijing Key Laboratory of Advanced Nuclear Materials and Physics,  Beihang
 University, Beijing 102206, China}
 \affiliation{School of Physics and Microelectronics,
Zhengzhou University, Zhengzhou, Henan 450001, China}

\author{Benjam\'in Grinstein}
\affiliation{Department of Physics, University of California, San Diego, La Jolla, California, 92093, USA}

\author{Sebastian J{\"a}ger}
\affiliation{Department of Physics and Astronomy, University of Sussex, Brighton BN1 9QH, United Kingdom}

\author{Shuang-Yi Li}
\affiliation{School of Physics,  Beihang University, Beijing 102206, China}

\author{Jorge Martin Camalich}
\affiliation{Instituto de Astrofisica de Canarias, C/ Via Lactea, s/n E38205 - La Laguna (Tenerife), Spain}
\affiliation{Universidad de La Laguna, Departamento de Astrofisica,
  La Laguna, Tenerife E-38205, Spain}
  
  \author{Rui-Xiang Shi}
\affiliation{School of Physics,  Beihang University, Beijing 102206, China}

\begin{abstract}
Motivated by renewed evidence for new physics in $b \to s\ell\ell$ transitions  in the form of LHCb's new measurements of theoretically
clean lepton-universality ratios and the purely leptonic $B_s\to\mu^+\mu^-$ decay, we quantify the combined level of discrepancy with the Standard Model and fit values of short-distance
Wilson coefficients. A combination of the clean observables $R_K$, $R_{K^*}$, and $B_s\to \mu\mu$ alone results in a discrepancy with the Standard Model at $4.0\sigma$, up from $3.5\sigma$ in 2017.
One-parameter scenarios with purely left-handed or with purely axial coupling to muons fit the data well and result in a $\sim 5 \sigma$ pull from 
the Standard Model. In a two-parameter fit of 
new-physics contributions with both vector and axial-vector couplings to muons the
allowed region is much more restricted than in 2017, principally due to the much more precise
result on $B_s \to \mu^+ \mu^-$, which probes the axial coupling to muons.
Including angular observables data restricts the allowed region further.
A by-product of our analysis is an updated average of
$\text{BR}(B_s \to \mu^+ \mu^-) = (2.8\pm 0.3) \times 10^{-9}$.
\end{abstract}

\maketitle

\section{Introduction}

Flavor physics played a central role in the development of the Standard Model (SM) and could well spearhead the discovery of new physics (NP) beyond the SM (BSM). In fact, although the vast majority of particle-physics data is consistent with the predictions of the SM,
a conspicuous series of discrepancies has appeared in rare
flavor-changing processes mediated by quark-level $b\to s\ell\ell$ transitions. These are suppressed by the ``GIM mechanism'' in the SM and are, therefore, potentially sensitive to very high-energy NP scales~\cite{Buras:2020xsm}. A perennial question in
this context is how to distinguish long-distance strong-interaction
effects from genuine new physics. Several years ago, following
LHCb's first measurement of the lepton-universality violating 
ratio $R_{K^*}$, we demonstrated
\cite{Geng:2017svp} the power of using observables which
are almost entirely free from hadronic uncertainties to
provide a high-significance rejection of the SM, and its potential
to narrow down the chiral structure of the BSM effect. In particular,
we pointed out the importance of the $B_s \to \mu^+ \mu^-$ decay
and lepton-flavor-violating ratios of forward-backward asymmetries
in lifting a degeneracy between axial and vectorial couplings to
leptons. Motivated by LHCb's updates to the ratio $R_K$ and
of $\text{BR}(B_s \to \mu^+ \mu^-)$ we revisit this set of decays in
the present work.

\section{Observables}
\label{sec:obs}
The measurement of several rare $b \to s \ell \ell$ decays yields results in tension with the SM expectations implying the presence of new interactions breaking lepton universality (see Refs.~\cite{Bifani:2018zmi,Buras:2020xsm} for recent reviews). Among them stand out a subset of observables with
theoretical uncertainties at or below the percent level~\cite{Hiller:2003js,Beneke:2017vpq}. Our main observables of interest comprise the lepton-universality ratios $R_K=\Gamma(B\to K\mu\mu)/\Gamma(B\to Kee)$~\cite{Hiller:2003js} and
$R_{K^*}=\Gamma(B\to {K^*}\mu\mu)/\Gamma(B\to {K^*}ee)$, and the
purely leptonic decay $B_s \to \mu^+ \mu^-$.

In particular, the LHCb collaboration has just reported the most precise measurement of $R_K$ in the $q^2$-bin $[1.1,6]~{\rm GeV}^2$ using the full run 1 and 2 datasets~\cite{Aaij:2021vac},
\begin{eqnarray}
R_K=0.846^{+0.042+0.013}_{-0.039-0.012}\;,
\end{eqnarray}
where the first uncertainty is statistical and the second systematic. This result deviates from the SM predictions~(see Table I in Ref.~\cite{Geng:2017svp})~\footnote{This prediction does not include the effect of electromagnetic corrections, which are of the order of a few percent~\cite{Bordone:2016gaq,Isidori:2020acz}. Experiments  subtract these effects; even if this subtraction was imperfect the resulting percent-level error is at present negligible
in light of the statistical uncertainties.}
\begin{eqnarray}
R_K^{\rm SM}=1.0004^{+0.0008}_{-0.0007}\;,
\end{eqnarray}
with a significance of $3.1\sigma$. Compared to the first LHCb  measurement reported in 2014~\cite{Aaij:2014ora}, the tension with respect to the SM has significantly increased.

At the same time, LHCb has published new results for the branching faction of $B_s\to\mu^+\mu^-$, 
\begin{eqnarray}
{\rm BR}(B_s\to\mu^+\mu^-)=(3.09^{+0.46+0.15}_{-0.43-0.11}) \times 10^{-9},
\end{eqnarray}
obtained with the same full dataset~\cite{Santimaria} and also known to about 1\% accuracy in the SM. 
This result, along with other recent measurements done by ATLAS~\cite{Aaboud:2018mst} and CMS~\cite{Sirunyan:2019xdu}, indicate a decay rate lower than the SM prediction. This set of key inputs is summarized in Table~\ref{tab:data}. In place of the asymmetric errors on $R_K$ and $R_{K^*}$ published by the experiments,
we conservatively employ a symmetric error equal to the upper, larger error (combining statistical and systematic in quadrature), in line with the treatment in  Ref.~\cite{Geng:2017svp}.

\begin{table*}[htb]
    \centering
\caption{Key inputs used in this paper.}
\label{tab:data}
{\normalsize
  \begin{tabular}{cccc}
\hline\hline
Observable & Value & Source & Reference\\
\hline
\multirow{5}{*}{ ${\rm BR}(B_s \to \mu^+ \mu^-)$} & $ (2.8 {}^{+0.8}_{-0.7}) \times 10^{-9}$ & ATLAS &\cite{Aaboud:2018mst} \\
 & $(2.9 \pm 0.7 \pm 0.2) \times 10^{-9}$ & CMS &\cite{Sirunyan:2019xdu} \\
 & $(3.09^{+0.46+0.15}_{-0.43-0.11}) \times 10^{-9}$ & LHCb update &  \cite{Santimaria} \\
 & $(2.842 \pm 0.333) \times 10^{-9}$ & Our average & This work \\
& $(3.63 \pm 0.13) \times 10^{-9}$ & SM prediction & \cite{Beneke:2019slt} \\
\hline
$R_K[1.1 , 6]$ & $0.846\pm0.044$& LHCb & \cite{Aaij:2021vac} \\
$R_K[1 , 6]$ & $1.03\pm0.28$& Belle & \cite{Abdesselam:2019lab} \\
\hline
$R_{K^*}[0.045,1.1]$ & $0.660\pm0.113$ & LHCb & \cite{Aaij:2017vbb} \\
$R_{K^*}[1.1,6]$ & $0.685\pm0.122$& LHCb & \cite{Aaij:2017vbb}\\
$R_{K^*}[0.045,1.1]$ & $0.52\pm0.365$ & Belle & \cite{Abdesselam:2019wac} \\
$R_{K^*}[1.1,6]$ & $0.96\pm0.463$& Belle & \cite{Abdesselam:2019wac}\\

\hline\hline
\end{tabular}
}
\end{table*}

In 2020 LHCb also reported a new measurement of the $CP$-averaged angular observables of the decay $B^0\to K^{*0}\mu^+\mu^-$~\cite{Aaij:2020nrf} and of its isospin partner, $B^+\to K^{*+}\mu^+\mu^-$~\cite{Aaij:2020ruw}. The new data seems to confirm the previous measurements pointing to possible tensions with the SM~\cite{DAmico:2017mtc,Ciuchini:2020gvn,Bhutta:2020qve,Becirevic:2020ssj,Alasfar:2020mne,Hurth:2020rzx,Bhom:2020lmk,Descotes-Genon:2020buf,
Biswas:2020uaq,Bordone:2019uzc,Coy:2019rfr,Bhattacharya:2019dot,
Arbey:2019duh,Kowalska:2019ley,Alguero:2019ptt,Aebischer:2019mlg,Ciuchini:2019usw,Alok:2019ufo,Alguero:2019pjc,Kumar:2019qbv,Datta:2019zca}. However, and contrary to the lepton-universality ratios and $B_s\to\mu\mu$, the SM predictions for the $B\to K^*\mu\mu$ angular observables suffer from significant hadronic uncertainties which hinder a clear interpretation of the discrepancies in terms of NP~\cite{Beneke:2001at,Grinstein:2004vb,Egede:2008uy,Khodjamirian:2010vf,Beylich:2011aq,Khodjamirian:2012rm,DescotesGenon:2012zf,Jager:2012uw,Horgan:2013pva,Lyon:2014hpa,Descotes-Genon:2014uoa,Jager:2014rwa,Straub:2015ica,Ciuchini:2015qxb,Hiller:2017bzc,Chobanova:2017ghn,Bobeth:2017vxj,Aaij:2016cbx,Gubernari:2018wyi}. 

In this work we combine the experimental data focusing on the clean observables as in Ref.~\cite{Geng:2017svp} and carry out global fits of the Wilson coefficients (or short-distance coefficients) of the low-energy $b\to s\ell\ell$ effective Lagrangian to the data.
 We find that the data on clean observables is at variance with the SM at a level of 4.0$\sigma$. We also find that one-parameter scenarios with purely left-handed or axial currents provide a good description of the data, excluding the SM point in each case at close to 5$\sigma$. As discussed abundantly in the literature, such new lepton-universality-violating
 (LUV) interactions can arise at
tree or loop level from new mediators such as neutral vector bosons ($Z'$) or leptoquarks (see Ref.~\cite{Cerri:2018ypt} which includes a review of NP interpretations). 

\subsection{Combination of ${\rm BR}(B_s \to \mu^+ \mu^-$) data}
\begin{figure}[h!]
\begin{tabular}{cc}
  \includegraphics[width=8.5cm]{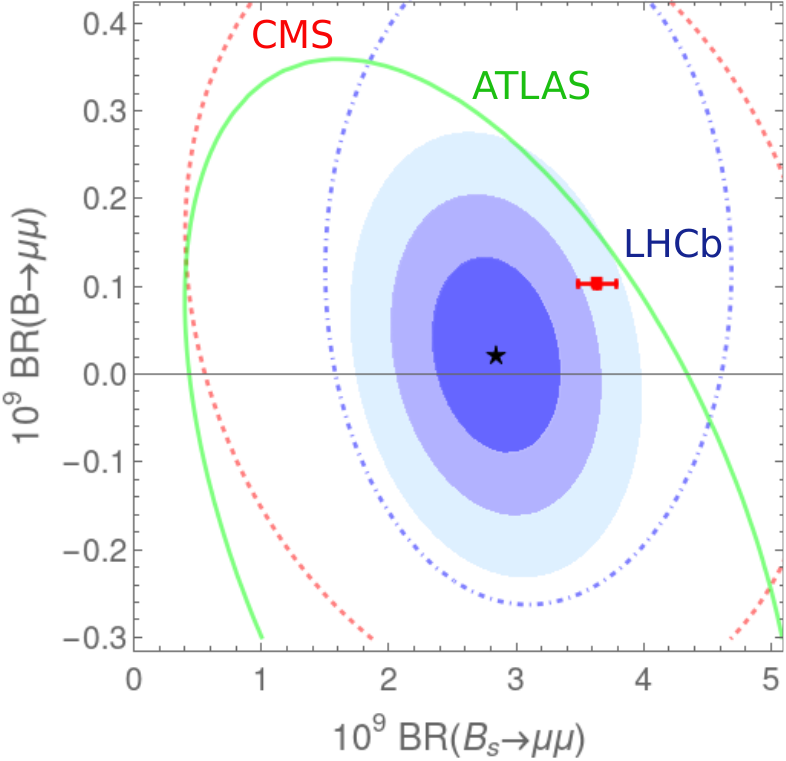}
\end{tabular}
\caption{Our combination of measurements of $\text{BR}(B_{s,d} \to \mu^+ \mu^-$) by ATLAS \cite{Aaboud:2018mst}, CMS \cite{Sirunyan:2019xdu}, and LHCb  \cite{Santimaria}, compared to the SM prediction (red square). Contours of the combination correspond to 1$\sigma$, 2$\sigma$ and 3$\sigma$, and those of each experiment to just 3$\sigma$.\label{Fig:Bqmumu}}
\end{figure}

An important aspect to note is that the three measurements of
$B_s \to \mu^+ \mu^-$ cannot be naively averaged together, as a result
of correlations with $B_d \to \mu^+ \mu^-$. We therefore
construct a two-dimensional joint likelihood from the published
measurements \cite{Aaboud:2018mst,Sirunyan:2019xdu,Santimaria}. In doing so, we assume a correlation coefficient of $-0.5$
for ATLAS, which reproduces the results reported in Ref.~\cite{Aaboud:2018mst}, and neglect correlations in the LHCb
measurement. The resulting combination is represented in
Fig.~\ref{Fig:Bqmumu}. Profiling over $\text{BR}(B_d \to \mu^+ \mu^-)$ results in
\begin{eqnarray}
\label{eq:Bsmumu}
\text{BR}(B_s \to \mu^+ \mu^-) &=& (2.8 \pm 0.3) \times 10^{-9}
\end{eqnarray}
with $\chi^2_{\rm min} = 3.72$ (5 d.o.f.).~\footnote{As usual, we treat the d.o.f. as the difference of the total number of data and fitted parameters. For example, in the case of $B_q\to \mu^+\mu^-$ we have six data points and one parameter to fit, ${\rm BR}(B_s\to\mu^+\mu^-)_{\rm exp}$.}
As with the existing combination~\cite{LHCb:2020zud}, the central value of the average
is lower than the average of the three individual central values. 

We combine the experimental measurements and the SM prediction of the $B_s\to\mu^+\mu^-$ branching fraction in the ratio 
 \begin{eqnarray}
 R=\frac{{\rm BR}(B_s^0\to\mu^+\mu^-)_{\rm exp}}{{\rm BR}(B_s^0\to\mu^+\mu^-)_{\rm SM}},
 \end{eqnarray}
obtaining $R=0.78(9)$ by using the most up to date
theoretical prediction of Ref.~\cite{Beneke:2017vpq}. We end this section by noting that only the recent LHCb result implements a newer (and larger by $\sim 6\%$) measurement of the ratio of hadronization fractions $f_s/f_d$. However, including the corresponding increase in the branching fractions of $B_s\to\mu^+\mu^-$ measured by ATLAS and CMS (keeping the correlation with $B_d\to\mu^+\mu^-$) leads to a very small increase (of about $\sim3\%$) in the average in Eq.~\eqref{eq:Bsmumu} that will be neglected in this work.

\section{Theoretical approach}

The low-energy effective Hamiltonian for semileptonic $b\rightarrow s\ell^+\ell^-$ processes at the scale $\mu\sim m_b$ in the SM is written as~\cite{Buchalla:1995vs}
\begin{eqnarray}
{\cal{H}}_{\rm eff}^{\rm SM}=\frac{4G_F}{\sqrt{2}}\sum_{p=u,c}\lambda_{ps}\left(C_1{\cal O}_1^p+C_2{\cal O}_2^p+\sum_{i=3}^{10}C_i{\cal O}_i\right),
\end{eqnarray}
where $G_F$ is the Fermi constant, and $\lambda_{ps}=V_{pb}V_{ps}^*$ is a combination of Cabibbo-Kobayashi-Maskawa (CKM) matrix elements with $p=u,c$. The short-distance contributions, stemming from scales above $\mu\sim m_b$, are matched to a set of Wilson coefficients $C_i$. The ${\cal O}_{1,2}^p$, ${\cal O}_{3-6}$ and ${\cal O}_8$ are the ``current-current'', ``QCD-penguin'' and ``chromomagnetic'' operators, respectively. The explicit forms for these operators can be found in Ref.~\cite{Buchalla:1995vs}. The remaining operators ${\cal O}_7$, ${\cal O}_9$, and ${\cal O}_{10}$ from electromagnetic penguin-, electroweak penguin-, and box-loop diagrams are defined as follows:
\begin{eqnarray}
&&\qquad\qquad\qquad O_7=\frac{e}{16\pi^2}m_b(\bar{s}\sigma^{\mu\nu}P_Rb)F_{\mu\nu},\nonumber\\
&&O_{9}^\ell=\frac{e^2}{16\pi^2}(\bar{s}\gamma^{\mu}P_Lb)(\bar{\ell}\gamma_{\mu}\ell),~~
O_{10}^\ell=\frac{e^2}{16\pi^2}(\bar{s}\gamma^{\mu}P_Lb)(\bar{\ell}\gamma_{\mu}\gamma_5\ell).\nonumber\\
\end{eqnarray}

In the presence of NP, one has nine more operators, i.e., $O_7'$ and $O_{9,10}'^\ell$, the
opposite-quark-chirality counterparts of $O_7$ and $O_{9,10}^\ell$, plus four scalar and two tensor operators~\cite{Alonso:2014csa}. However, if the NP effect enters in the couplings to muons, only  ${\cal O}_9$ and ${\cal O}_{10}$ can  explain the $R_{K^{(*)}}$ data~\cite{Alonso:2014csa}. The tensor-operator contributions to $b\to s\ell\ell$ decays are suppressed if the NP scale is heavier than the electroweak scale. Other operators cannot induce LUV, or are tightly constrained by the $B_q\to\ell\ell$ decays. In the following analysis,  we assume that all the Wilson coefficients are real and  that the presence of NP only appears in the $b\to s\mu^+\mu^-$ sector. The last assumption is justified by the tension in the data in $b\to s\mu^+\mu^-$, in particular in BR($B_s\to\mu^+\mu^-$) discussed above.

For reliable predictions of observables, it is of vital importance to estimate theoretical uncertainties. They mainly stem from nonperturbative contributions including form factors $F(q^2)$'s and  ``nonfactorizable'' terms $h_\lambda(q^2)$'s~\cite{Beneke:2001at}. At low $q^2$ the nonperturbative contributions can be addressed in the heavy quark and large-energy limits: they can be  expanded as~\cite{Jager:2014rwa},
\begin{eqnarray}
&&F(q^2)=F^\infty(q^2)+a_F+b_Fq^2/m_B^2+{\cal O}\left((q^2/m_B^2)^2\right),\nonumber\\
&&h_\lambda(q^2)=h_\lambda^\infty(q^2)+r_\lambda(q^2),
\end{eqnarray}
where  $F^\infty(q^2)$ and $h_\lambda^\infty(q^2)$ can be calculated in light-cone sum rules~\cite{Ball:2004rg,Khodjamirian:2010vf}  and within the QCD factorization approach~\cite{Beneke:2001at}, and  the rest are power-correction terms. Among them,  $r_\lambda(q^2)$ is dominated by the long-distance charm contributions involving the ``current-current'' operators ${\cal O}_1^c$ and ${\cal O}_2^c$~\cite{Jager:2014rwa}. We parametrize   the charm loop contributions, $r_\lambda^c(q^2)$, by
\begin{eqnarray}
r_\lambda^c(q^2)=A_\lambda+B_\lambda\frac{q^2}{4m_c^2},
\end{eqnarray}
where $A_\lambda$ and $B_\lambda$ are dimensionless constants. Therefore, the overall uncertainties at low $q^2$ arise from the leading-power terms, power corrections of form factors and charm loop contributions, characterized by 27 nuisance parameters. For more details and ranges of values taken for these, see Refs.~\cite{Jager:2014rwa,Geng:2017svp}. In the high-$q^2$ region, form factors have been calculated in lattice QCD~\cite{Horgan:2013pva} whereas the nonfactorizable contribution can be computed with an operator product expansion~\cite{Grinstein:2004vb,Beylich:2011aq}. We  omit the analysis of this region as none of the LUV ratios have been measured there yet. 

We use the frequentist statistical approach to quantify the compatibility between the experimental data and the theoretical predictions. We define the $\chi^2$ function as
\begin{eqnarray}
\tilde \chi^2(\vec C,~\vec y)=\chi^2_{\rm exp}(\vec C,~\vec y)+\chi^2_{\rm th}(\vec y),
\end{eqnarray}
where $\chi^2_{\rm exp}(\vec C,~\vec y)$   includes the correlations reported by the experiments and $\chi^2_{\rm th}(\vec y)$ is a theoretical component.

The theoretical predictions for the observables $O^{\rm th}$ are functions of Wilson coefficients $\vec C$ and nuisance hadronic parameters $\vec y$. We choose two models for $\chi^2_{\rm th}(\vec y)$; one in which $y_i$ follows a normal distribution  (that we call ``Gaussian'') and another (that we call ``$R$-fit'') where it is restricted to a range, see Ref.~\cite{Jager:2014rwa}, with  a flat distribution. Here, we assume that these nuisance parameters are uncorrelated~\cite{Jager:2014rwa}.

In order to obtain best-fit values in a particular scenario, we can construct a profile $\chi^2$ depending only on certain Wilson coefficients
\begin{eqnarray}
\chi^2(\vec C)=\underset{\vec{y}}{\rm min}\,\tilde\chi^2(\vec{C},\vec{y}),
\end{eqnarray}
with the remaining Wilson coefficients set to their SM values. Here, $\chi^2$ is minimized by varying the nuisance parameters.
In our statistical analysis, we adopt  the widely used $p$-value and ${\rm Pull}_{\rm SM}$ to denote how well the experimental data can be described and how significant is the deviation from the SM. Results of fits are reported for $\delta C_i\equiv C_i-C_I^{\text{SM}}$.

\begin{table*}[htb]
\caption{Best-fit values, $\chi_{\rm min}^2$, $p$-value, ${\rm Pull}_{\rm SM}$ and confidence intervals of the Wilson coefficients in the fits of the $R_K$, $R_{K^*}$, $B_s\to \mu\mu$ data only using Gaussian form $\chi^2_{\rm th}$. For the cases of single Wilson-coefficient fits, we show the $1\sigma$ and $3\sigma$ confidence intervals. In the $(\delta C_9^{\mu},\delta C_{10}^{\rm \mu})$ case, the $1\sigma$ interval of each Wilson coefficient is obtained by profiling over the other one to take into account their correlation.}
\label{tab:GuassianRKRKsBstouuFit}
{\normalsize
  \begin{tabular}{cccccccc}
\hline\hline
Coefficient & Best fit & $\chi^2_{\rm min}$ & $p$-value & ${\rm Pull}_{\rm SM}$ [$\sigma$] & 1$\sigma$ range & 3$\sigma$ range & $\rho$\\
\hline
$\delta C_9^{\mu}$ & $-0.82$  & 14.70 [6 d.o.f.] & 0.02 & 4.08 &$[-1.06,-0.60]$  & $[-1.60,~-0.20]$ & $\cdots$\\

$\delta C_{10}^{\mu}$ & 0.65  & 6.52 ~~[6 d.o.f.] & 0.37 & 4.98 & $[0.52,0.80]$ & $[0.25,1.11]$ & $\cdots$\\

$\delta C_L^{\mu}$ & $-0.40$ & 7.36 [6 d.o.f.] & 0.29 & 4.89 & $[-0.48,-0.31]$  & $[-0.66,-0.15]$ & $\cdots$\\
\hline
$(\delta C_9^{\mu},\delta C_{10}^{\rm \mu})$ & $(-0.11, 0.59)$ & \multirow{2}{*}{6.38~~[5 d.o.f.]} & \multirow{2}{*}{0.27} & \multirow{2}{*}{4.62}
   & $\delta C_9^\mu \in$ $[-0.41,~0.17]$ & $\delta C_{10}^\mu \in$ $[0.38,~0.81]$ & $0.762$\\
$(\delta C_L^{\mu},\delta C_{R}^{\rm \mu})$    & $(-0.35, 0.25)$&&&&$\delta C_L^{\mu}\in [-0.45,-0.26]$&$\delta C_R^{\mu}\in [0.00,0.48]$& $0.406$\\
\hline\hline
\end{tabular}
}
\end{table*}

\section{The theoretically clean fit}

We first restrict ourselves to the analysis of the theoretically clean observables  $R_K$, $R_{K^*}$ and $B_s\to \mu^+\mu^-$. The relevant data is shown in Table~\ref{tab:data} and discussed above. We first assess the consistency of the dataset, which gives $\chi^2_{\rm min}= 4.61$ ($8$ d.o.f.), corresponding to $p=0.80$, where $p$ denotes $p$-value. For the d.o.f. we count the 6 $B_q \to \mu^+ \mu^-$ measurements separately and for $R_K$ we count the LHCb and Belle results as two separate measurements of the same observable, neglecting the small difference in the lower end of the bin.

\begin{figure}[h!]
\begin{tabular}{cc}
  \includegraphics[width=8.5cm]{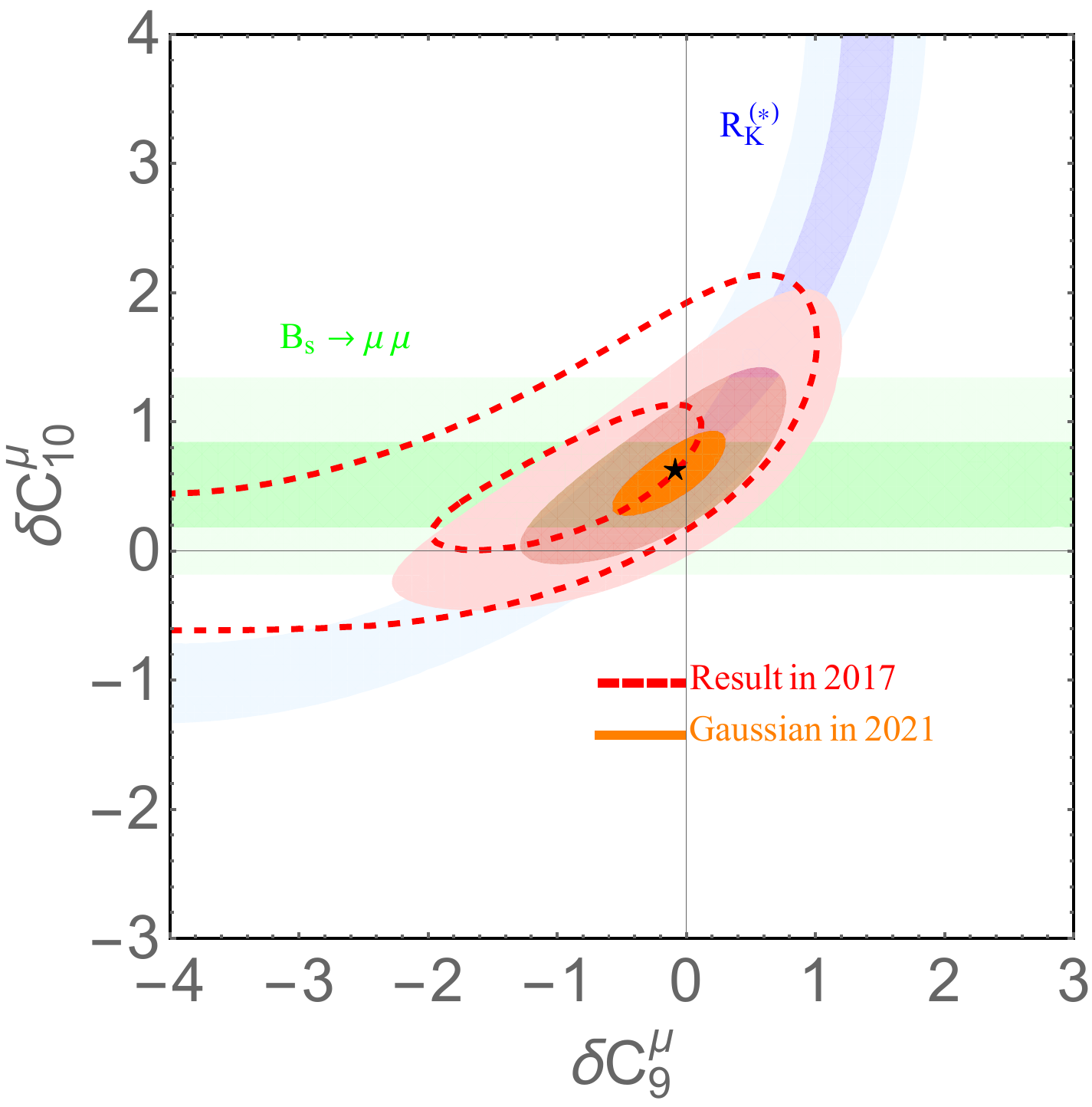}
\end{tabular}
\caption{Contour plots at the $1\sigma$, $3\sigma$ and $5\sigma$ confidence levels in the $(\delta C_9^{\mu},\delta C_{10}^{\mu})$ plane for the Gaussian form of $\chi^2_{\rm th}$ (regions in light red and orange). We also show the $1\sigma$ and $3\sigma$ contours for the individual constraints ($R_{K^{(*)}}$ and $B_s\to\mu\mu$) and, for the same fit from 2017~\cite{Geng:2017svp} (dashed  in red).  
\label{Fig:RKRKsBsuuFit}}
\end{figure}

Minimizing $\chi^2$ over all SM and theoretical nuisance parameters, one obtains $\chi_{\rm min,SM}^2=31.32$ ($\chi_{\rm min,SM}^2=30.54$) and $p_{\rm SM}=5.4\times10^{-5}$ ($p_{\rm SM}=7.6\times10^{-5}$) using the Gaussian ($R$-fit) form of the $\chi^2_{\rm th}$. We emphasize that the sole role of the minimization here is to implement the tiny theoretical uncertainties. In arriving at this $p$-value and in the rest of this paper, we treat our ${\rm BR}(B_s \to \mu^+ \mu^-)$ average as a single measurement, following
common practice and in line with Ref.~\cite{Geng:2017svp}. In other words, with
7 d.o.f. the clean data is at variance with the
null hypothesis (Standard Model) at a level of $4.0\sigma$,
up from $3.5\sigma$ in our previous work \cite{Geng:2017svp}. 
Were we  to treat the 6 $B_q \to \mu^+ \mu^-$ measurements as separate inputs,
we would instead obtain $\chi^2_{\rm min, SM} = 35.04$ and $p_{\rm SM} = 3.5\sigma$ (12 d.o.f.);
the reduction in significance is unsurprising given that we now include (nonanomalous)
data on $b \to d \ell \ell$ transitions.
We next fit one- and two-parameter BSM scenarios. Only $\delta C_9^\mu$ and $\delta C^\mu_{10}$ can describe a deficit in both $R_K$ and $R_{K^*}$~(Ref.~\cite{Geng:2017svp} and Fig.~1 there). Therefore we  analyze the data in the clean observables by fitting only these two Wilson coefficients.
 Figure~\ref{Fig:RKRKsBsuuFit} shows the constraints imposed in the $(\delta C_9^\mu, \delta C_{10}^\mu)$ plane and using the Gaussian model for $\chi^2_{\rm th}$ (see below for the $R$-fit results). In Table~\ref{tab:AllobsFit} we also show the numerical results for this fit, as well as fits involving a single Wilson coefficient, for the Gaussian approach. We also fit the left- and right-handed combinations (related to the couplings to muons) of Wilson coefficients $\delta C_L^{\mu}=(\delta C_9^{\mu}-\delta C_{10}^{\mu})/2$ and $\delta C_R^{\mu}=(\delta C_9^{\mu}+\delta C_{10}^{\mu})/2$. We observe that one-parameter
scenarios with purely left-handed coupling $\delta C_L^\mu$ or purely axial coupling $\delta C_{10}^\mu$ describe the data well ($p>10\%$). Compared to either scenario, the SM (identified
as the $\delta C_{i}=0$ point) is excluded at a confidence level of close to $5\sigma$. On the other hand, and in contrast with our previous analysis from 2017~\cite{Geng:2017svp}, the pure-$C_9$ scenario is in tension with the data at $2.3\sigma$ although it still compares favorably with the data compared to the SM. Finally, as we will see below, using the $R$-fit version of $\chi^2_{\rm th}$ and increasing the theoretical uncertainties in the predictions of $B\to K^{(*)}\ell\ell$ only produce very small changes in the results  of the fit to clean observables and do not change our conclusions.
 
\begin{figure}[h!]
  \includegraphics[width=8.5cm]{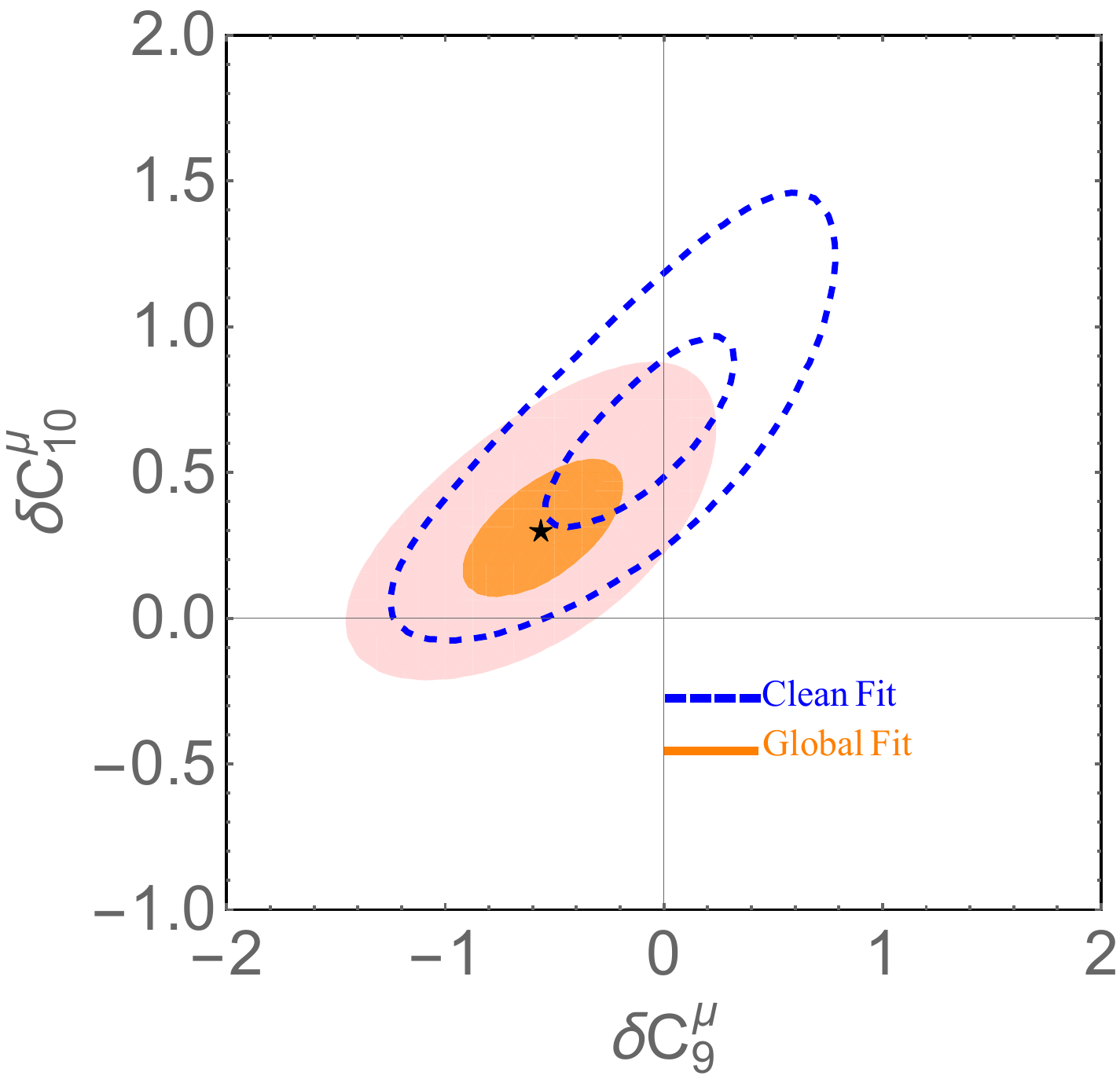}\\
  \caption{Contour plots at the $1\sigma$ and $3\sigma$ confidence level for the $(\delta C_9^{\mu},\delta C_{10}^{\mu})$ scenario. For comparison, we show the global fits with all observables for Gaussian~(regions in light red and orange) and the corresponding clean fits of Fig.~2~(dashed lines in blue)\label{Fig:AllobsFit}.}
\end{figure}

\section{The global fit}

\begin{table*}[htb]
    \centering
\caption{The same as Table~\ref{tab:GuassianRKRKsBstouuFit} but fitting to the $R_K$, $R_{K^*}$, $B_s\to \mu\mu$ and angular observables of $B\to K^*\mu\mu$ data.}
\label{tab:AllobsFit}
{\normalsize
  \begin{tabular}{cccccccc}
\hline
\hline
Coefficient & Best fit & $\chi^2_{\rm min}$ & ~~~$p$-value~~~ & ~~~${\rm Pull}_{\rm SM}$~~~ & 1$\sigma$ range & 3$\sigma$ range & $\rho$\\
\hline
\hline
$\delta C_9^{\mu}$ & $-0.85$ & 106.32 [93 d.o.f.] & 0.16 & 4.53 & $[-1.06, -0.64]$ & $[-1.50, -0.27]$ & $\cdots$\\
$\delta C_{10}^{\mu}$ & 0.54 & 107.82 [93 d.o.f.] & 0.14 & 4.37 & [0.41, 0.67] & [0.16, 0.94] & $\cdots$\\
$\delta C_L^{\mu}$ & $-0.39$ & 102.81 [93 d.o.f.] & 0.23 & 4.91 & $[-0.48, -0.31]$ & $[-0.65, -0.15]$ & $\cdots$\\

\hline
$(\delta C_9^{\mu},\delta C_{10}^{\mu})$ & $(-0.56,0.30)$ & \multirow{2}{*}{102.36 [92 d.o.f.]} & \multirow{2}{*}{0.22} & \multirow{2}{*}{4.58} & $\delta C_9^{\mu}\in$ $[-0.79, -0.31]$ & $\delta C_{10}^{\mu}\in[0.15, 0.49]$ & $0.317$ \\
$(\delta C_L^{\mu},\delta C_{R}^{\mu})$&$(-0.43,-0.12)$&&&&$\delta C_L^{\mu}\in[-0.52,-0.33]$&$\delta C_R^{\mu}\in[-0.27,0.03]$& $0.364$\\
\hline
\hline
\end{tabular}
}
\end{table*}

For the sake of completeness we also perform a global fit including all the measurements of angular observables reported by the LHCb, ATLAS, and CMS experiments in the low-$q^2$ region. As mentioned above, these observables are afflicted by larger theoretical uncertainties compared to LUV ratios and $B_s\to\mu\mu$. However, it is important to analyze how the conclusions change when including these data within a model-independent framework for the theoretical uncertainties such as ours. 

More specifically, compared to our 2017 analysis~\cite{Geng:2017svp}, we replace
the $CP$-averaged angular observables for the $B^0\to K^{*0}\mu^+\mu^-$, the ratio $R$ for the $B_s\to\mu^+\mu^-$ decay, and $R_K$ in the bin~$[1.1,~6.0]~{\rm GeV}^2$ with the latest measurements by the LHCb, CMS and ATLAS experiments~\cite{Aaij:2019wad,Aaij:2020nrf,Sirunyan:2019xdu,Aaboud:2018mst,LHCb:2020zud}. In addition, we also include 32 new measurements of $F_L$, $P_1$, $P_2$, $P_3$, $P_4'$, $P_5'$, $P_6'$ and $P_8'$ in four low bins~($q^2\leq6~{\rm GeV}^2$) for the $B^+\to K^{*+}\mu^+\mu^-$ decay~\cite{Aaij:2020ruw} as well as three Belle data $R_K$ and $R_{K^*}$ in Table~\ref{tab:data}. As a result the total number of data fitted becomes 94.~\footnote{We note that the total number of  data fitted in Ref.~\cite{Geng:2017svp} is 59, not 65, because we used the ATLAS measurements~\cite{Aaboud:2018krd} in the wide bin~$[0.04,~4.0]~{\rm GeV}^2$ for the $CP$-averaged angular observables not two separate bins~$[0.04,~2.0]~{\rm GeV}^2$ and ~$[2.0,~4.0]~{\rm GeV}^2$. This only affected the computed $p$-value. }

 In this fit strategy, we obtain a $\chi_{\rm min,SM}^2=126.88$ and $p_{\rm SM}=0.01$ with 94 d.o.f.. Compared to the global fit results in Ref.~\cite{Geng:2017svp}, the updated fit results in Table~\ref{tab:AllobsFit} and Fig.~\ref{Fig:AllobsFit} show that the confidence level of the exclusion of the SM point increases by $1.3\sigma$ for the $\delta C_{10}^{\mu}$ scenario and by $0.7\sigma$ for the $\delta C_L^{\mu}$ scenario, but only by $0.2\sigma$ and $0.4\sigma$ in the $\delta C_9^{\mu}$ and $(\delta C_9^{\mu},\delta C_{10}^{\mu})$ scenarios, respectively. Interestingly, these updated fits constrain better these Wilson coefficients and exclude positive $\delta C_9^\mu$ and negative $\delta C_{10}^\mu$ at more than the $3\sigma$ confidence level. 

We also perform a four-dimensional global fit with the Gaussian $\chi^2_{\rm th}$ including $C_9^{\prime\mu}$ and $C_{10}^{\prime\mu}$. The resulting Wilson coefficients from the fit are,
\begin{eqnarray}\label{eq:globalfit_results}
 \left(
\begin{array}{c}
\delta C_9^\mu\\
\delta C_{10}^\mu\\
 C_9^{\prime\mu}\\
 C_{10}^{\prime\mu}
\end{array}
\right)=\left(
\begin{array}{c}
-1.07\pm0.29\\
0.21\pm0.14\\
0.32\pm0.21\\
-0.26\pm0.14
\end{array}
\right),
\end{eqnarray}
with the correlation matrix,
\begin{eqnarray}\label{eq:globalfit_rho}
 \rho=\left(
\begin{array}{cccc}
 1.000 & 0.529 &-0.381 & 0.455\\
  & 1.000 & 0.010 & 0.263\\
  & & 1.000 & 0.153\\
 &  & & 1.000
\end{array}
\right),
\end{eqnarray}
and where $\chi_{\rm min}^2=96.88$ for 90 d.o.f., corresponding to a $p$-value of 0.29 and a Pull$_{\rm SM}=4.57$.

\section{Impact of theoretical uncertainties}

\begin{figure}[h!]
  \includegraphics[width=8.5cm]{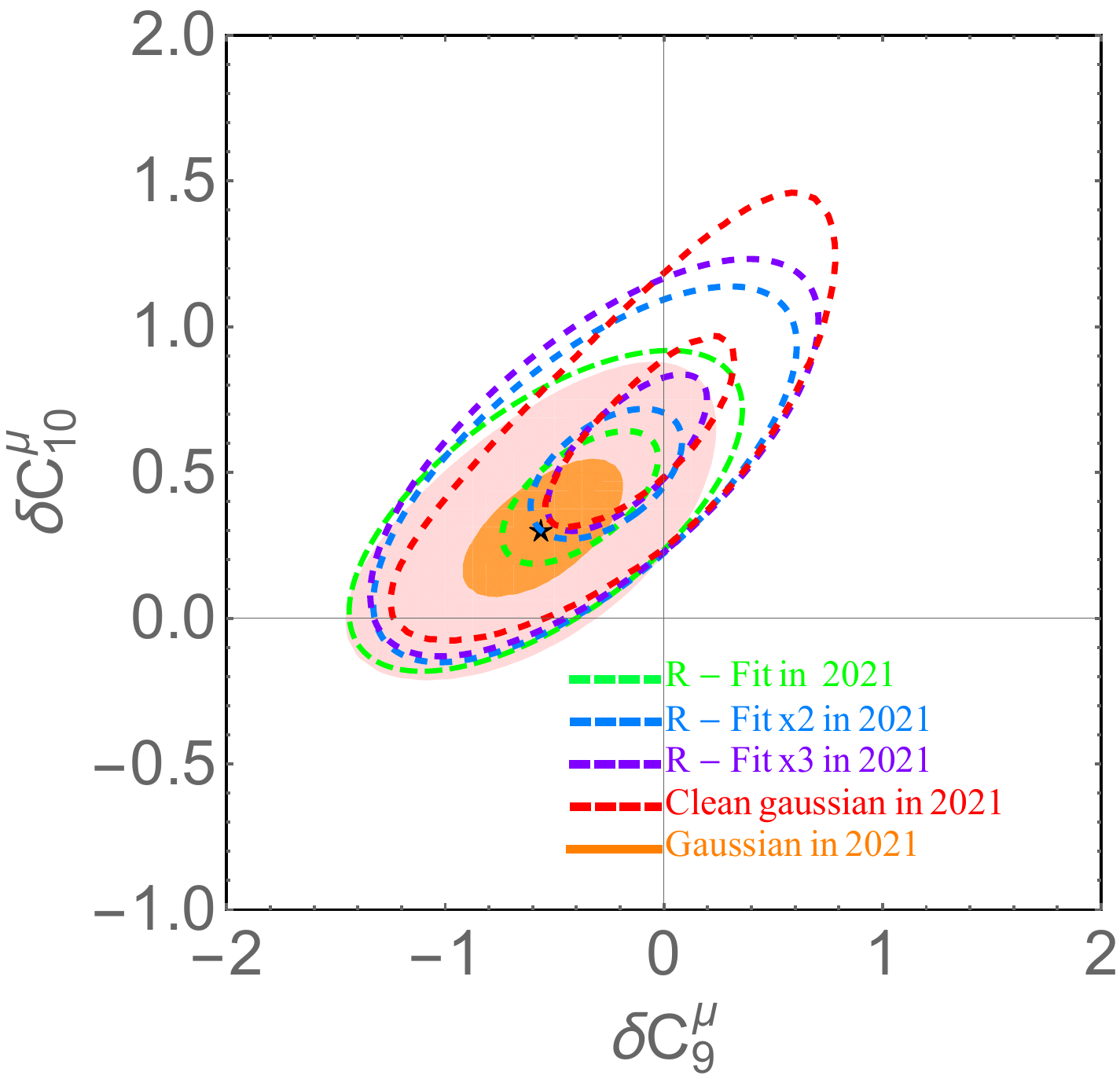}\\
  \caption{Contour plots at the $1\sigma$ and $3\sigma$ confidence level for the $(\delta C_9^{\mu},\delta C_{10}^{\mu})$ scenario in the global fit. For comparison, we show the clean fit (dashed lines in red) in the Gaussian method, and global fits with all observables using the Gaussian~(regions in light red and orange) and $R$-fit methods with nominal hadronic uncertainties in $B\to K^{(*)}\ell\ell$~(dashed lines in green), or multiplied by 2~(dashed lines in blue) and 3~(dashed lines in purple).\label{Fig:AllobsRFit}}
\end{figure}

Finally, we briefly investigate the robustness of the fits with respect to the hadronic uncertainties. We do so by comparing the results obtained above with those obtained by using the $R$-fit model for $\chi^2_{\rm th}$, with nominal hadronic uncertainties in $B\to K^{(*)}\ell\ell$ or multiplied by a factor 2 and 3. The relevant results are shown in Tables~\ref{tab:AllobsRFit}, \ref{tab:AllobsRFitx2} and ~\ref{tab:AllobsRFitx3}. In Fig.~\ref{Fig:AllobsRFit} we also show the new results in the $(\delta C_9^\mu,~\delta C_{10}^\mu)$ plane overlaid with the ones obtained with the Gaussian model of $\chi^2_{\rm th}$. 

The treatment of uncertainties has a significant impact on the global fit, especially in the parameter ranges obtained for the Wilson coefficient $\delta C_9^\mu$. As discussed in Refs.~\cite{Jager:2012uw,Jager:2014rwa} this is due to the fact that a shift to
$C_9$ in the amplitude is indistinguishable from a nonfactorizable
charm contribution or a shift to a certain combination of $B$-decay form factors. Therefore, increasing the ranges allowed for these parameters in a framework such as $R$-fit tends to reduce the significance of a NP effect in $C_9$. 

\begin{figure}[h!]
  \includegraphics[width=8.5cm]{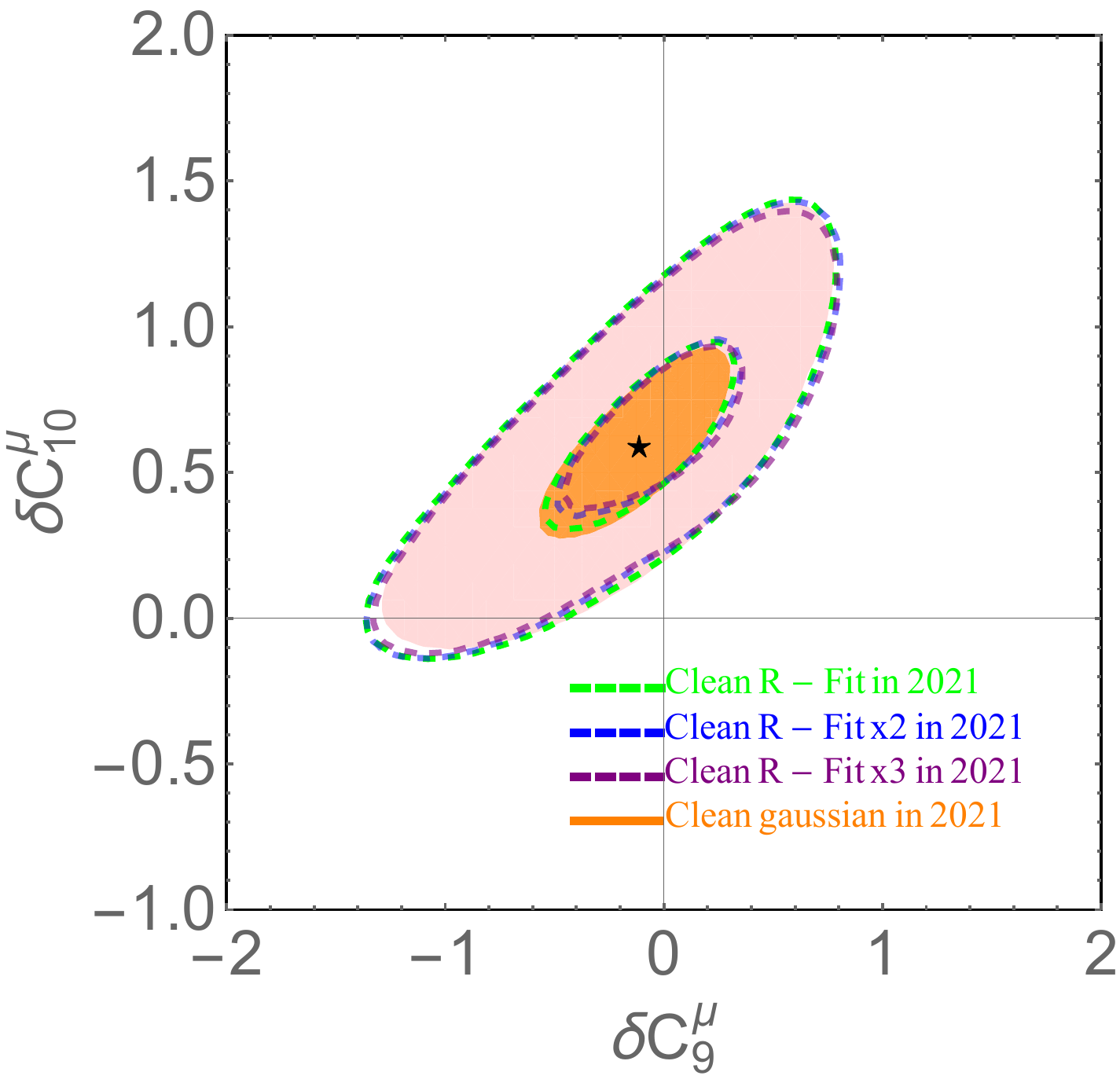}\\
  \caption{Contour plots at the $1\sigma$ and $3\sigma$ confidence level for the $(\delta C_9^{\mu},\delta C_{10}^{\mu})$ scenario in the clean fit. For comparison, we show the clean fits using the Gaussian~(regions in light red and orange) and $R$-fit methods with nominal hadronic uncertainties in $B\to K^{(*)}\ell\ell$ (dashed lines in green, corresponding to $\chi^2_{\rm SM,min}=30.54$ and $p$-value$_{\rm SM}=7.6\times10^{-5}$), or multiplied by 2~(dashed lines in blue, corresponding to $\chi^2_{\rm SM,min}=29.99$ and $p$-value$_{\rm SM}=9.5\times10^{-5}$) and 3~(dashed lines in purple, corresponding to $\chi^2_{\rm SM,min}=29.54$ and $p$-value$_{\rm SM}=1.15\times10^{-4}$).\label{Fig:CleanobsRFit}}
\end{figure}

This effect is clearly seen in Fig.~\ref{Fig:AllobsRFit} where the contours in the global fit approach those of the clean fit when increasing the errors and the tension of the data with the SM becomes dominated by the LUV ratios and $B_s\to \mu\mu$. In contrast,  the results and conclusions derived from the clean fit to these latter observables are robust with respect to the same variation of hadronic uncertainties. This is illustrated in Fig.~\ref{Fig:CleanobsRFit}.


\begin{table*}[htb]
    \centering
\caption{Same as Table~\ref{tab:AllobsFit} but for the $R$-fit method with nominal hadronic uncertainties as given in Ref.~\cite{Geng:2017svp}. One obtains $\chi^2_{\rm min,SM}=121.19$, corresponding to a $p$-value of 0.03.}
\label{tab:AllobsRFit}
{\normalsize
  \begin{tabular}{ccccccc}
\hline
\hline
Coefficient & Best fit & $\chi^2_{\rm min}$ & ~~~$p$-value~~~ & ~~~${\rm Pull}_{\rm SM}$~~~ & 1$\sigma$ range & 3$\sigma$ range\\\cline{1-5}
\hline
$\delta C_9^{\mu}$ & $-0.86$ & 102.3 [93 d.o.f.] & 0.24 & 4.35 & $[-1.10, -0.69]$ & $[-1.61, -0.24]$ \\
$\delta C_{10}^{\mu}$ & 0.56 & 99.24 [93 d.o.f.] & 0.31 & 4.69 & [0.45, 0.67] & [0.24, 0.96] \\
$\delta C_L^{\mu}$ & $-0.40$ & 96.32 [93 d.o.f.] & 0.39 & 4.99 & $[-0.48, -0.32]$ & $[-0.64, -0.16]$ \\

\hline

$(\delta C_9^{\mu},\delta C_{10}^{\mu})$ & $(-0.51,0.36)$ & 96.17 [92 d.o.f.] & 0.36 & 4.63 & $\delta C_9^{\mu}\in$ $[-0.63, -0.19]$ & $\delta C_{10}^{\mu}\in$[0.24, 0.54] \\\cline{1-5}
\hline
\hline
\end{tabular}
}
\end{table*}
\begin{table*}[htb]
    \centering
\caption{Same as Table~\ref{tab:AllobsFit} but for the $R$-fit method with hadronic uncertainties of $B\to K^{(*)}\ell\ell$ multiplied by 2. One obtains $\chi^2_{\rm min,SM}=117.18$, corresponding to a $p$-value of 0.05.}
\label{tab:AllobsRFitx2}
{\normalsize
  \begin{tabular}{ccccccc}
\hline
\hline
Coefficient & Best fit & $\chi^2_{\rm min}$ & ~~~$p$-value~~~ & ~~~${\rm Pull}_{\rm SM}$~~~ & 1$\sigma$ range & 3$\sigma$ range\\\cline{1-5}
\hline
$\delta C_9^{\mu}$ & $-0.88$ & 99.95 [93 d.o.f.] & 0.29 & 4.15 & $[-1.11, -0.68]$ & $[-1.63, -0.22]$ \\
$\delta C_{10}^{\mu}$ & 0.58 & 93.18 [93 d.o.f.] & 0.48 & 4.90 & [0.48, 0.66] & [0.23, 0.99] \\
$\delta C_L^{\mu}$ & $-0.40$ & 92.90 [93 d.o.f.] & 0.48 & 4.93 & $[-0.48, -0.32]$ & $[-0.64, -0.16]$ \\

\hline

$(\delta C_9^{\mu},\delta C_{10}^{\mu})$ & $(-0.19,0.60)$ & 92.19 [92 d.o.f.] & 0.47 & 4.63 & $\delta C_9^{\mu}\in$ $[-0.53, 0.00]$ & $\delta C_{10}^{\mu}\in$[0.32, 0.71] \\\cline{1-5}
\hline
\hline
\end{tabular}
}
\end{table*}
\begin{table*}[htb]
    \centering
\caption{Same as Table~\ref{tab:AllobsFit} but for the $R$-fit method with hadronic uncertainties of $B\to K^{(*)}\ell\ell$ multiplied by 3. One obtains $\chi^2_{\rm min,SM}=115.03$, corresponding to a $p$-value of 0.07.}
\label{tab:AllobsRFitx3}
{\normalsize
  \begin{tabular}{ccccccc}
\hline
\hline
Coeff. & best fit & $\chi^2_{\rm min}$ & ~~~$p$-value~~~ & ~~~${\rm Pull}_{\rm SM}$~~~ & 1$\sigma$ range & 3$\sigma$ range\\\cline{1-5}
\hline
$\delta C_9^{\mu}$ & $-0.86$ & 97.52 [93 d.o.f.] & 0.35 & 4.18 & $[-1.10, -0.68]$ & $[-1.64, -0.23]$ \\
$\delta C_{10}^{\mu}$ & 0.70 & 89.40 [93 d.o.f.] & 0.59 & 5.06 & [0.61, 0.81] & [0.27, 1.02] \\
$\delta C_L^{\mu}$ & $-0.41$ & 90.27 [93 d.o.f.] & 0.56 & 4.98 & $[-0.49, -0.33]$ & $[-0.65, -0.17]$ \\

\hline

$(\delta C_9^{\mu},\delta C_{10}^{\mu})$ & $(0.02,0.70)$ & 89.37 [92 d.o.f.] & 0.56 & 4.69 & $\delta C_9^{\mu}\in$ $[-0.28, 0.13]$ & $\delta C_{10}^{\mu}\in$[0.39, 0.82] \\\cline{1-5}
\hline
\hline
\end{tabular}
}
\end{table*}

\section{Summary and outlook}

In conclusion, we have presented a statistical analysis of recent data on LUV ratios and $B_s\to\mu\mu$  using the low-energy $b\to s\ell\ell$ effective Lagrangian. We  find  that  the  data  on these clean observables disagree with the SM at a level of $4.0\sigma$. Scenarios with pure left-handed or axial currents provide a good description of the data, and each of them excludes the SM point at $\sim 5\sigma$ confidence level. Therefore, our results reinforce the NP interpretation of the anomalies in the $b\to s\ell\ell$ transitions, which could correspond to the contribution of the tree-level exchange of a leptoquark or $Z^\prime$ with a mass $\Lambda\sim 30$~TeV and $\sim\mathcal O(1)$ couplings to the SM. Further data on LUV observables from LHCb and  Belle II should be able to clarify soon this tantalizing possibility. 

\section{Acknowledgments}
 We would like to thank A.\ Cerri, M.\ Bona and R.\ Zwicky for
helpful conversations and P.\ Hern\'andez for useful comments.
This work is partly supported by the National Natural Science Foundation of China under Grants No.~11735003, No.~11975041, and No.11961141004, the Academic Excellence  Foundation of BUAA for Ph.D. students, and the fundamental Research Funds
for the Central Universities.  The work of B.G. is supported in part by the U.S. Department of Energy Grant No. DE-SC0009919. S.J. is supported in part by the U.K. Science and Technology Facilities Council under Consolidated Grants No.~ST/P000819/1 and No.~ST/T00102X/1. J.M.C. acknowledges support from the Spanish MINECO through the ``Ram\'on y Cajal'' Program No.~RYC-2016-20672 and Grant No.~PGC2018-102016-A-I00.

$Note~added.$-Recently, several other papers appeared performing similar global fits and finding very similar conclusions~\cite{Angelescu:2021lln,Altmannshofer:2021qrr,Cornella:2021sby,Kriewald:2021hfc,Lancierini:2021sdf,Hurth:2021nsi,Alguero:2021anc}.

\clearpage
\section{Appendix: Updated clean fits after the latest LHCb measurements of $R_{K^0_s}$ and $R_{K^{*+}}$ }

In this section, we show that the latest measurements of the branching fraction ratios $R_{K^0_s}$ and $R_{K^{*+}}$, in combination with the already known ratios  $R_{K^+}$, $R_{K^{*0}}$, and the branching fraction $B_s\to \mu\mu$,  point at a discrepancy with the Standard Model at $4.2\sigma$.
One-parameter scenarios, $C_{10}^\mu$ and $ C_L^\mu$, fit the data well and result in a  pull from
the SM crossing the 5.0$\sigma$ threshold. The  two-parameter fit of $C_9$ and $C_{10}$ yields also a pull of 5.02$\sigma$. On the other hand, the one-parameter scenario of $C_9$ alone still has a pull of 4.5$\sigma$, up from the previous 4.1$\sigma$ (see Table II).

Very recently, the LHCb Collaboration reported  measurements of two new lepton-universality ratios $R_{K_S^0}=\Gamma(B^0\to K_S^0\mu^+\mu^-)/\Gamma(B^0\to K_S^0 e^+ e^-)$ and $R_{K^{*+}}=\Gamma(B^+\to K^{*+}\mu^+\mu^-)/\Gamma(B^+\to K^{*+} e^+ e^-)$ in the $q^2$ ranges $[1.1,6.0]~{\rm GeV}^2$ and $[0.045,6.0]~{\rm GeV}^2$, respectively, using proton-proton collision data corresponding to an integrated luminosity of 9 fb$^{-1}$~\cite{LHCb:2021lvy}. They represent the first observation of the  $B^0\to K_s^0 e^+ e^-$ and $B^+\to K^{*+} e^+ e^-$ decays. The two ratios are
\begin{eqnarray}
&&R_{K_S^0}=0.66_{~-0.14~-0.04}^{~+0.20~+0.02},\nonumber\\
&&R_{K^{*+}}=0.70_{~-0.13~-0.04}^{~+0.18~+0.03},
\end{eqnarray}
where the first error is statistical and the second is systematic. These results show again tension with respect to the SM predictions (see Table I) with a significance of $1.5\sigma$ and $1.4\sigma$, respectively. In the following analysis, we conservatively employ a symmetric error equal to the larger error.

It should be noted that these ratios are the isospin partners of $R_{K^+}$ and $R_{K^{*0}}$, and therefore should  receive the same NP contributions, if they exist. As a result, it is of utmost importance to update the clean fit performed in the main text and to check whether these new measurements increase or decrease the significance of the tension with the SM.

Note that compared to the clean fit performed in the main text, the total number of fitted data becomes 9 after adding the two new LHCb measurements. Setting the Wilson coefficients to their SM values, we obtain a $\chi^2_{\rm min,SM}=36.50$, corresponding to a $p$-value of $3.23\times10^{-5}$ or a $4.2\sigma$ deviation (for 9 d.o.f.). As a result,  with the new data, the LU ratios are in discrepancy with the SM predictions at a level of $4.2\sigma$, up from the $4.0\sigma$ in March 2021.

\begin{figure}[h!]
\centering
\begin{tabular}{cc}
  \includegraphics[width=8.5cm]{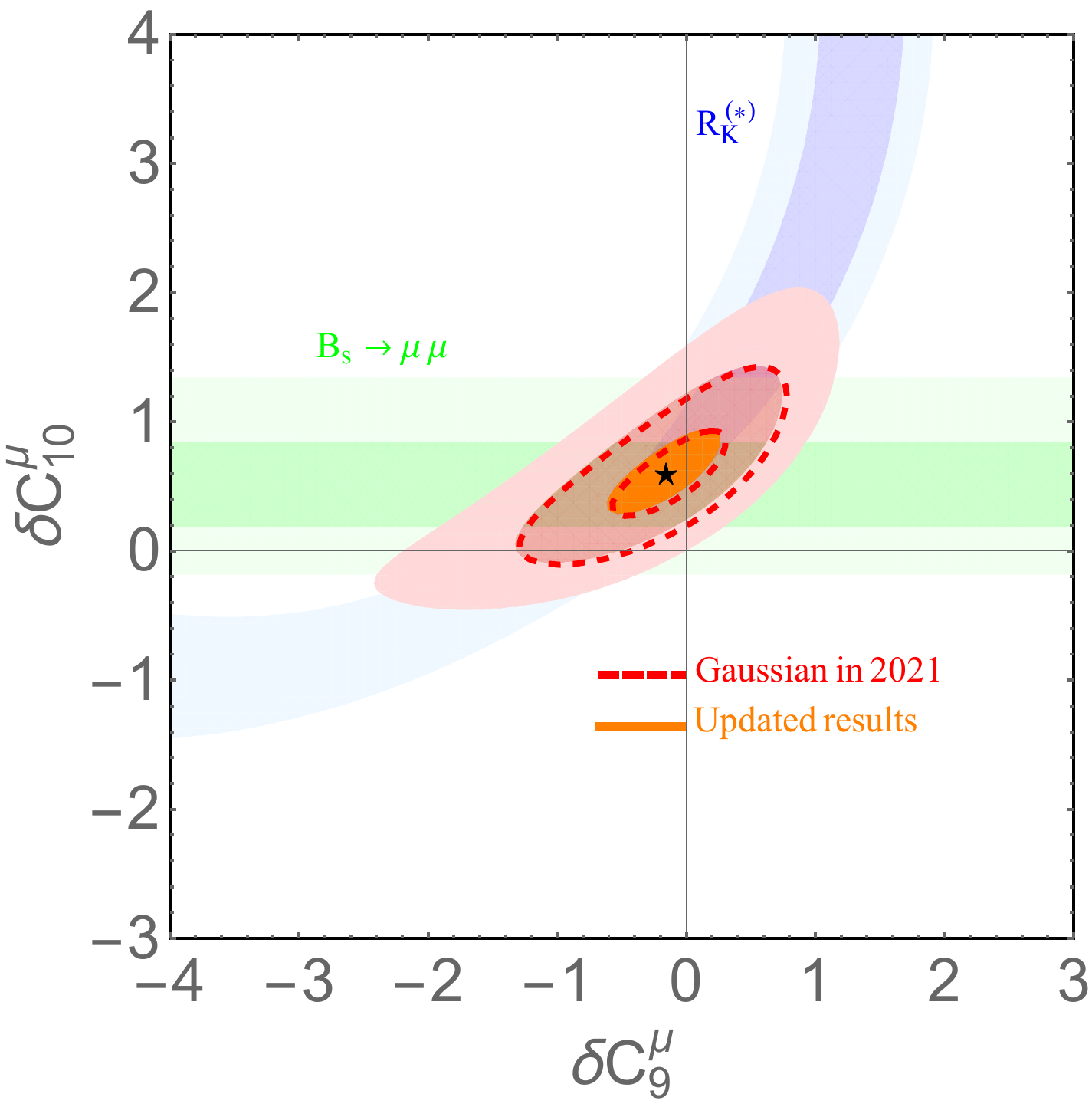}
\end{tabular}
\caption{Same as Fig.~\ref{Fig:RKRKsBsuuFit} but for the new clean fit to 9 data including the latest LHCb measurements~\cite{LHCb:2021lvy} vs. the clean fit to 7 data shown in Fig.~\ref{Fig:RKRKsBsuuFit} (Gaussian in 2021).
\label{updatedFig:Cleanfitplot}}
\end{figure}

In Table~\ref{updatedtab:Cleangauss} and Fig.~\ref{updatedFig:Cleanfitplot}, we show the results of four fits which are obtained by allowing for
lepton-specific contributions $\delta C_9^\mu$, $\delta C_{10}^\mu$, $\delta C_L^\mu$ and a two-dimensional scenario $(\delta C_9^\mu, \delta C_{10}^\mu)$. We find that the significance of all the NP scenarios except for $\delta C_9^\mu$ are more than $5\sigma$. Such a result, obtained by considering only clean observables, seems to point unambiguously to the presence of new physics. We conclude that although the newly observed ratios are the isospin partners of the existing ones,  they indeed have increased the significance of the tension with the SM by about $0.5\sigma$ compared with the fits without them in March 2021.

\begin{table*}[htb]
\caption{Same as Table~\ref{tab:GuassianRKRKsBstouuFit} but fitted also to the latest LHCb data $R_{K_S^0}$ and $R_{K^{*+}}$.}
\label{updatedtab:Cleangauss}
{\normalsize
  \begin{tabular}{cccccccc}
\hline\hline
Coefficient & Best fit & $\chi^2_{\rm min}$ & $p$-value & ${\rm Pull}_{\rm SM}$ & 1$\sigma$ range & 3$\sigma$ range & $\rho$\\
\hline
$\delta C_9^{\mu}$ & $-0.89$  & 16.27 [8 d.o.f.] & 0.04 & 4.50 &$[-1.12,-0.67]$  & $[-1.67,~-0.27]$ & $\cdots$\\

$\delta C_{10}^{\mu}$ & 0.69  & 7.84 ~~[8 d.o.f.] & 0.45 & 5.35 & $[0.57,0.84]$ & $[0.29,1.16]$ & $\cdots$\\

$\delta C_L^{\mu}$ & $-0.42$ & 8.45 [8 d.o.f.] & 0.39 & 5.30 & $[-0.51,-0.34]$  & $[-0.69,-0.18]$ & $\cdots$\\
\hline
$(\delta C_9^{\mu},\delta C_{10}^{\rm \mu})$ & $(-0.15, 0.60)$ & $7.56$~[7 d.o.f.] & $0.37$ & $5.02$
   & $\delta C_9^\mu \in$ $[-0.45,~0.13]$ & $\delta C_{10}^\mu \in$ $[0.10,~1.01]$ & $0.765$\\
\hline\hline
\end{tabular}
}
\end{table*}

\bibliography{RKs}
\end{document}